\documentclass[preprint]{aastex62}
\usepackage{amsmath,amstext}
\usepackage{graphicx}

\def\vec#1{\ensuremath{\mathbf{#1}}}

\newcounter{RomanNumber}
\newcommand{\MyRoman}[1]{\setcounter{RomanNumber}{#1}\Roman{RomanNumber}}

\received{****}
\revised{****}
\accepted{****}
\shorttitle{Non-Equilibrium Ionization in forward models of sausage modes}
\shortauthors{Shi et al.}
\begin{document}

\title{Non-equilibrium Ionization Effects on Extreme-Ultraviolet Emissions
    Modulated by Standing Sausage Modes in Coronal Loops}

\correspondingauthor{Bo Li}
\email{bbl@sdu.edu.cn}

\author{Mijie Shi}
\affiliation{Shandong Provincial Key Laboratory of Optical Astronomy and Solar-Terrestrial Environment, Institute of Space Sciences, Shandong University, Weihai 264209, China}
\affiliation{CAS Key Laboratory of Solar Activity, National Astronomical Observatories, Beijing 100012, China}
\author{Bo Li}
\affiliation{Shandong Provincial Key Laboratory of Optical Astronomy and Solar-Terrestrial Environment, Institute of Space Sciences, Shandong University, Weihai 264209, China}

\author{Tom Van Doorsselaere}
\affiliation{Centre for Mathematical Plasma Astrophysics, Department of Mathematics, Celestijnenlaan 200B, B-3001, KU~Leuven, Leuven, Belgium}

\author{Shao-Xia Chen}
\affiliation{Shandong Provincial Key Laboratory of Optical Astronomy and Solar-Terrestrial Environment, Institute of Space Sciences, Shandong University, Weihai 264209, China}

\author{Zhenghua Huang}
\affiliation{Shandong Provincial Key Laboratory of Optical Astronomy and Solar-Terrestrial Environment, Institute of Space Sciences, Shandong University, Weihai 264209, China}



\begin{abstract}
Forward-modeling the emission properties in various passbands is important
    for confidently identifying magnetohydrodynamic waves in the structured solar corona.
We examine how Non-equilibrium Ionization (NEI) affects
    the Extreme Ultraviolet (EUV) emissions modulated by standing fast sausage modes (FSMs)
    in coronal loops, taking the Fe \MyRoman{9} 171 \AA\ and
	Fe \MyRoman{12} 193 \AA\ emission lines as examples.
Starting with the expressions for linear FSMs in straight cylinders,
    we synthesize the specific intensities and spectral profiles for the two spectral lines
    by incorporating the self-consistently derived ionic fractions in the relevant contribution functions.
We find that relative to the case where Equilibrium Ionization (EI) is assumed,
    NEI considerably impacts the intensity modulations,
    but shows essentially no effect on the Doppler velocities or widths.
Furthermore, NEI may affect the phase difference between intensity variations and those
    in Doppler widths for Fe \MyRoman{12} 193 \AA\ when the line-of-sight
    is oblique to the loop axis.
While this difference is $180^\circ$ when EI is assumed, it is $\sim 90^\circ$
    when NEI is incorporated for the parameters we choose.
We conclude that in addition to viewing angles and instrumental resolutions,
    NEI further complicates the detection of
    FSMs in spectroscopic measurements of coronal loops
    in the EUV passband.
\end{abstract}




\section{Introduction} 
\label{sec:intro}
The past two decades have witnessed rapid developments of coronal seismology,
    thanks to the abundantly identified low-frequency waves and oscillations
    in the highly structured solar corona
    \citep[see recent reviews by e.g.,][]
    {2005LRSP....2....3N, 2007SoPh..246....3B, 2012RSPTA.370.3193D, 2016SSRv..200...75N}.
However, identifying a measured oscillatory signal with a specific magnetohydrodynamic (MHD) wave mode
    is not straightforward.
Take the deceptively simple case of slow waves in coronal loops, and assume that the spatial dependence of the
    associated perturbations is restricted to the axial direction.
Starting with an analytical model for the fluid parameters,
    the forward-modeling effort by \citet[][and references therein]{2008SoPh..252..101D}
    demonstrated that the periods in the modulated intensities in, say, Fe \MyRoman{12} 195\AA,
    do not necessarily correspond to the wave period, let alone the damping rates.
This results from the intricate dependence of the emissivity on density and ionization balance.
(See also \citeauthor{2016ApJ...825...58R}~\citeyear{2016ApJ...825...58R}
    for a more recent forward-modeling study on slow waves in the corona.)
The situation becomes even trickier when one considers the distribution of wave perturbations transverse to
    coronal loops, because this further complicates the integration
    of emissivities along a Line-of-Sight (LoS).
Fast sausage modes (FSMs) are the simplest in this regard because they are axisymmetric and hence avoid
    the additional complication associated with the azimuthal dependence
    \citep[see e.g.,][for more discussions]{2016ApJS..223...23Y, 2017ApJ...836..219A}.
A variety of forward-modeling studies have been conducted with different levels of sophistication,
    starting from the works by
    \citet{2003A&A...397..765C} and \citet{2003A&A...409..325C}
	who computed the modulated intensities by integrating squared densities along LoS
    with different viewing angles.
This approach was taken further by \citet{2012A&A...543A..12G}
	who examined the effects of spatial resolution, namely the ``width" of an LoS.
Further incorporating the contribution function into the computations by assuming
    equilibrium ionization (EI),
	\citet[hereafter AvD13]{2013A&A...555A..74A} derived the spectral profiles
    of Fe \MyRoman{9}~171~\AA\ and Fe \MyRoman{12}~193~\AA\ emission lines.
As found from this series of studies, the observability of FSMs in EUV emissions
    depends rather sensitively on such geometrical parameters as viewing angles,
    and on instrumental parameters like temporal and spatial resolutions as well.

Similar to AvD13, this study will also address the spectral properties
    of Fe \MyRoman{9} 171 \AA\ and Fe \MyRoman{12} 193 \AA\ as modulated
    by standing FSMs.
New is that non-equilibrium ionization (NEI) is addressed when computing
    the ionic fractions of Fe \MyRoman{9} and~\MyRoman{12}.
The reason for doing this is that the periods of FSMs in coronal loops
    are determined by the transverse
    Alfv\'en time, which typically attains a couple of seconds
    \citep[e.g.,][]{1970A&A.....9..159R,1975IGAFS..37....3Z,1982SoPh...75....3S,1983SoPh...88..179E,
    1986SoPh..103..277C}.
However, the ionization and recombination timescales for the relevant
    ionization states are comparable to, or even substantially longer than the
    wave period in the case of Fe \MyRoman{12} (see Table 1 in AvD13).
This means that in general the ionic fractions cannot respond instantaneously to the variations
    in the electron temperature, and differences from the EI
    computations are expected.
To isolate the effects of NEI, we will examine the simplest configuration where
    FSMs are hosted by a straight, axially homogeneous cylinder with
    physical parameters distributed in a piece-wise constant manner transverse
    to the cylinder.
By doing this, we are avoiding the complications due to the continuous
    transverse structuring \citep[e.g.,][]{2012ApJ...761..134N, 2014A&A...568A..31L, 2015ApJ...812...22C,
    2016ApJ...833..114C, 2018JPhA...51b5501C}.
In addition, we will focus only on trapped modes such that no apparent attenuation
    is involved.
Section~\ref{sausage wave model} will formulate the FSMs and describe the equilibrium parameters,
    and Section~\ref{sec_descr_emission} will describe how the emission properties are computed.
We will present our results in Section~\ref{Results} before
    concluding this study in Section \ref{summary and conclusion}.

\section{LINEAR STANDING FAST SAUSAGE MODES IN CORONAL LOOPS}
\label{sausage wave model}

We model an equilibrium coronal loop as a static, straight cylinder
    with radius $R=1.5\times10^3$~km and length-to-radius ratio $L/R = 12.5$.
In a cylindrical coordinate system ($r, \phi, z$),
    both the cylinder axis and the equilibrium magnetic field ${\mathbf B}$
    are in the $z$-direction.
We adopt single-fluid ideal MHD and consider an electron-proton plasma throughout.
The equilibrium parameters are structured only in the $r$-direction,
    and subscript ${\rm i}$ (${\rm e}$) denotes the constant
    values inside (outside) the cylinder.
Let $N$, $T$, and $B$ denote the electron number density,
    electron temperature, and magnetic field strength, respectively.
We take $[N_{\rm i}, N_{\rm e}] = [5, 0.5] \times 10^9$~cm$^{-3}$,
    and $[T_{\rm i}, T_{\rm e}] = [1, 0.74]$~MK.
We further take $B_{\rm i} = 16.2$~G, and hence an external one
    $B_{\rm e} = 17.2$~G results from transverse force balance.
For reference, the internal (external) plasma beta is
    $\beta_{\rm i} = 0.13$ ($\beta_{\rm e} = 0.009$).
Furthermore, the Alfv\'en speed in the internal
    (external) medium reads $v_{\rm Ai} = 500$ ($v_{\rm Ae} = 1675$)~km~s$^{-1}$.

In this equilibrium, 
    standing linear FSMs perturb all physical parameters except the azimuthal
    components of the velocity and magnetic field.
Suppose that the system has reached a stationary state
    characterized by angular frequency $\omega$
    and axial wavenumber $k$.
The physical variables relevant for computing EUV emissions are given by
\begin{eqnarray}
&& N (r, z; t)    =
      N_0 [1-{\cal D}(r) \sin(\omega t)\sin(kz)]~,
         \label{eq_pert_den} \\[0.1cm]
&& v_r (r, z; t)  =
        \omega {\cal R}(r) \cos(\omega t)\sin(kz)~,
         \label{eq_pert_vr} \\[0.1cm]
&& \displaystyle 
   v_z (r, z; t)  =
      -\frac{c_{\rm s}^2}{\omega/k} {\cal D}(r) \cos(\omega t)\cos(kz)~,
         \label{eq_pert_vz} \\[0.1cm]
&& T (r, z; t)  =
      T_0 [1-(\gamma-1) {\cal D}(r) \sin(\omega t)\sin(kz)]~,
         \label{eq_pert_T}
\end{eqnarray}
    where both the equilibrium values (subscript $0$) and perturbations are involved,
    and $\gamma = 5/3$ is the adiabatic index.    
Here ${\cal R}$ denotes the transverse (i.e., radial) profile of the
    transverse Lagrangian displacement as given by
\begin{eqnarray}
 {\cal R} (r) = \left\{
   \begin{array}{ll}
   A R J_1(nr)~,	   				& r < R~, \\[0.2cm] 
   \displaystyle
   \frac{A R J_1(nR)}{K_1(mR)} K_1 (mr)~,	& r > R~, \\ 
   \end{array}\right.
\label{eq_trans_displ}
\end{eqnarray}
    where the constant $A$ determines the relative magnitude, and $J_1$ ($K_1$)
    is the Bessel function of the first kind (modified Bessel function of the second kind).
In addition, the effective transverse wavenumbers $n$ and $m$ are defined by
\begin{eqnarray*}
&& \displaystyle n^2 =  \frac{(\omega^2 - k^2 v_{\rm Ai}^2)(\omega^2 - k^2 c_{\rm si}^2)}
                             {(c_{\rm si}^2  +v_{\rm Ai}^2)(\omega^2 - k^2 c_{\rm Ti}^2)}~, \\[0.2cm]
&& \displaystyle m^2 = -\frac{(\omega^2 - k^2 v_{\rm Ae}^2)(\omega^2 - k^2 c_{\rm se}^2)}
                             {(c_{\rm se}^2  +v_{\rm Ae}^2)(\omega^2 - k^2 c_{\rm Te}^2)}~, 
\end{eqnarray*}
    where $c_{\rm s}$ and $c_{\rm T} = c_{\rm s} v_{\rm A}/\sqrt{c_{\rm s}^2 + v_{\rm A}^2}$
    are the adiabatic sound and tube speeds, respectively
\footnote{We use Bessel's $K$ function to describe the perturbations outside the tube because we will examine 
    trapped modes.
    In this case $\omega$ is real-valued, $n^2$ and $m^2$ are both non-negative
    \citep[see e.g.,][for more discussions on this aspect]{1986SoPh..103..277C, 
    2007AstL...33..706K,  2015ApJ...812...22C,2016ApJ...833..114C}.}.
For an electron-proton plasma, $c_s^2 = 2 \gamma k_{\rm B} T_0/m_{\rm p}$ with 
    $k_{\rm B}$ being the Boltzmann constant and $m_{\rm p}$ the proton mass. 
As for ${\cal D}$, it is related to ${\cal R}$ by
\begin{eqnarray}
 {\cal D} (r) = \frac{\omega^2}{\omega^2-k^2 c_{\rm s}^2}
                \frac{{\rm d} (r {\cal R})}{r {\rm d} r}~.
\label{eq_compressibity}
\end{eqnarray}
For future reference, we note that the Lagrangian displacements in the radial
    and axial directions are given by
\begin{eqnarray}
&& \xi_r (r, z ; t) = 
        {\cal R}(r) \sin(\omega t) \sin(kz)~, 
         \label{eq_xir} \\ 
&& \displaystyle 
   \xi_z (r, z ; t) = 
         -\frac{c_{\rm s}^2}{\omega^2/k} {\cal D}(r) \sin(\omega t)\cos(kz)~,
         \label{eq_xiz}
\end{eqnarray}
Finally, the angular frequency $\omega$ is found by solving the relevant
    dispersion relation~\citep[e.g.,][Eq.~8b]{1983SoPh...88..179E}.

We adopt the following parameters for the perturbations.
The axial wavenumber $k$ is taken to be $5 \pi/L$, corresponding to
    the fourth longitudinal harmonic.
Solving the dispersion relation then yields
    a wave period $P = 2\pi/\omega$ of $6.21$~secs
    for the transverse fundamental mode, which is in the trapped regime. 
Consequently, the axial phase speed $\omega/k$ reads 
    $2.41~v_{\rm Ai} = 1207$~km~s$^{-1}$.
The relative magnitude of the transverse displacement ($A$)
    is specified such that
    the peak value in $v_r$ is $0.06~v_{\rm Ai} = 30$~km~s$^{-1}$.
The transverse displacement $\xi_r$ can reach up to $0.02 R = 30$~km.        
The peak value in the perturbed density (temperature)
    reads $\sim 0.09N_i$ ($\sim 0.06T_i$).
As for the axial velocity, the peak value is only $\sim 0.0041~v_{\rm Ai}$
    ($2.07$~km~s$^{-1}$), which is readily understandable 
    because of the factor in front of ${\cal D}$ in 
    Equation~\eqref{eq_pert_vz}.

We now construct the spatial distributions of
    the fluid parameters in the $r-z$ plane with a spacing of $30$~km
    in both directions for $t$ between $0$ and $4$ periods.
While this is implemented on a Eulerian grid, 
    we take advection into account by assigning, to a point $\vec{x}$ at time $t$, 
    the physical parameters evaluated at $\vec{x}' = \vec{x}-\vec{\xi}(\vec{x}, t)$
    where $\vec{\xi}$ is the displacement vector
\footnote{
In principle, $\vec{x}'$ should be computed such that 
    $\vec{x}'+\vec{\xi}(\vec{x}', t) = \vec{x}$. 
But the difference between the two sets of values for $\vec{x}'$
    is $\vec{\xi}\cdot\nabla\vec{\xi}$ and therefore of
    second-order.}. 
Figure~\ref{fig:sausage_perturb} presents the spatial distributions
    in a cut, through the cylinder axis,
    of the fluid parameters at some representative instants of time.
The radial ($v_r$) and axial ($v_z$) speeds are shown for $t=0$,
    while the electron density ($N$) and temperature ($T$)
    are displayed for $t=P/4$.
From Figures~\ref{fig:sausage_perturb}c and \ref{fig:sausage_perturb}d,
    one can barely discern the expansion or contraction
    of the coronal tube. 
Different instants of time are chosen due to the $\pi/2$ phase difference between
    the relevant perturbations (see Equations~\ref{eq_pert_den}
    to \ref{eq_pert_T}).
In Figure~\ref{fig:sausage_perturb}a, the white dashed lines denote two lines-of-sight
    that both pass through the cylinder axis.
Let $\theta$ denote the angle between a LoS and the cylinder axis.
Then the LoS labeled 1 (2) corresponds to a $\theta$ being $0^\circ$ ($45^\circ$),
    chosen to represent normal (oblique) viewing angles that one frequently encounters
    in observations.

\section{COMPUTING EMISSION MODULATIONS DUE TO SAUSAGE MODES}
\label{sec_descr_emission}
The coupled equations governing the ionic fraction ($f_q$) for Fe of charge state $q$ 
   are given by
\begin{eqnarray}
\label{eq_ionic_frac}
\displaystyle
   \left(\frac{\partial}{\partial t}+{\mathbf v}\cdot\nabla\right) f_q
 = N \left[f_{q-1}C_{q-1}-f_q\left(C_q+R_q\right)+f_{q+1}R_{q+1}\right]~,
\end{eqnarray}
   where the ionization ($C$) and recombination ($R$) rate coefficients
   depend only on electron temperature $T$ and 
   are found with CHIANTI 
   (ver~8, 
   \citeauthor{2015A&A...582A..56D}~\citeyear{2015A&A...582A..56D})
   \footnote{http://www.chiantidatabase.org/}.
If neglecting the left-hand side (LHS), or equivalently assuming that 
   the wave period is much longer than the ionization timescales,
   then we end up with a set of coupled algebraic equations that pertain
   to EI, namely ionization balance.
By noting that there is no Fe $0$ or Fe \MyRoman{28}, one finds that the rate of 
   ionization into $q$ balances the rate of recombination out of $q$.
In other words,
\begin{eqnarray}
C_{q-1} f_{q-1} = R_{q} f_q~.
\label{eq_EI}
\end{eqnarray}
In reality, the wave period is not necessarily much longer than 
   the ionization timescales.
To take NEI into account, we then solve Equation~\eqref{eq_ionic_frac} 
   at each Eulerian grid point
   by initiating the solution procedure
   with the EI solution at time $t=0$.
The integration procedure is similar to earlier works~\citep[e.g.,][]{2010ApJ...722..625K, 2013ApJ...773..110S}
   in other contexts.
It suffices to consider only Fe \MyRoman{5} to \MyRoman{15},
   because the fractions of the rest of ionic states are negligible.
The time step for integrating Equation~\eqref{eq_ionic_frac}
   is small enough to resolve the ionization or recombination processes,
   and we make sure that $\left|\sum_{q={\mathrm{V}}}^{\mathrm{XV}}f_q-1\right|<10^{-5}$.

Figure~\ref{fig:ion_fraction} displays the temporal evolution of
   the ionic fractions of Fe \MyRoman{9} ($f_{\mathrm{IX}}$, Fig.~\ref{fig:ion_fraction}b)
   and Fe \MyRoman{12} ($f_{\mathrm{XII}}$, Fig.~\ref{fig:ion_fraction}c)
   at $[r, z] = [0, L/2]$, where the compressibility is the strongest. 	
In addition to the NEI results (the blue curves),
   their EI counterparts are also shown (red).
The temporal evolution of the electron temperature ($T$)
   at the same location is presented in Figure~\ref{fig:ion_fraction}a for reference.
As expected, the ionic fractions respond instantaneously to the variation in $T$
   in the equilibrium case.
In particular, one sees that
   $f_{\mathrm{IV}}$ ($f_{\mathrm{XII}}$) is
   in anti-phase (in-phase) with $T$ for the parameters considered.
This behavior can be understood as follows, for which purpose we define 
   $\Gamma_q = C_{q-1}/R_{q}$ and $\Pi_q = \Gamma_1 \Gamma_2 \cdots \Gamma_q$.
Note that $\Gamma_1$ is defined to be unity.
Then the algebraic equations pertaining to EI (Equation~\ref{eq_EI})
   yield that 
\begin{eqnarray}
 \displaystyle 
&&  f_{\rm I} = \frac{1}{\Pi_1 + \Pi_2 + \cdots + \Pi_{27}}~, 
    \label{eq_EI_f1} \\[0.2cm]
&&  f_{q} = f_{q-1} \Gamma_q = f_{\rm I} \Pi_{q} \hspace{0.5cm} \mbox{for } q\ge 2~.
    \label{eq_EI_fq}
\end{eqnarray}
In agreement with physical intuitions, $\Gamma_q$ decreases monotonically with $q$ at a fixed $T$
   and increases monotonically with $T$ at a fixed $q$.
And it turns out that either $\Gamma_{9}$ or $\Gamma_{10}$ is the last one that exceeds unity
   in the ${\Gamma_q}$ series
   in the examined temperature range.
\footnote{
Note that $\Gamma_1$ is defined to be unity for mathematical convenience, 
   and only the series $\{\Gamma_2, \Gamma_3, \cdots, \Gamma_{27}\}$ is physically relevant.
Note also that $\Gamma_q$ depends on both $q$ and the electron temperature ($T$).
Taking the perturbation due to the fast sausage mode 
   into account, we find that $T$ varies between 
   $0.736$ and $1.061$~MK.
In this range, $\Gamma_q$ ($q \le 9$)
   is consistently larger than unity, 
   whereas $\Gamma_q$ ($q \ge 11$)
   is consistently smaller than unity. 
However, $\Gamma_{10}$ exceeds unity only when $T \gtrsim 1.054$~MK.
Now that $\Gamma_q$ decreases with $q$ at a fixed $T$, one finds that 
   $\Gamma_9$ is the last element that exceeds unity in the $\{\Gamma_q\}$ series
   for $T \lesssim 1.054$~MK, and $\Gamma_{10}$
   takes up this role when $T \gtrsim 1.054$~MK.
For instance, when $T=1$~MK, one finds that 
   $[\Gamma_8, \Gamma_9, \Gamma_{10}, \Gamma_{11}]
   =[21.3, 6.71, 0.817, 0.411]$.
However, this series reads 
   $[24.4, 7.85, 1.03,  0.5251]$ when $T = 1.06$~MK.
}   
Regardless, the fraction of Fe \MyRoman{1} in EI can be expressed as
\begin{eqnarray}
 \displaystyle 
 f_{\rm I} = \frac{1/\Pi_9}{
  1+1/\Gamma_9
   +\Gamma_{10}+\Gamma_{10}\Gamma_{11}+\Gamma_{10}\Gamma_{11}\Gamma_{12}+\cdots
  }~,
 \label{eq_EI_f1_reduced}
\end{eqnarray}
  where we have neglected the terms represented by $\cdots$ because 
  they contribute no larger than $1\%$. 
Equation~\eqref{eq_EI_fq} then indicates that 
\begin{eqnarray}
 \displaystyle 
 f_{\rm IX} = \frac{1}{
  1+1/\Gamma_9
   +\Gamma_{10}+\Gamma_{10}\Gamma_{11}+\Gamma_{10}\Gamma_{11}\Gamma_{12}+\cdots
  }~.
 \label{eq_EI_f9_reduced}
\end{eqnarray}
Now define $' = {\rm d}/{\rm d}T$ and let $G_9$ denote the denominator. 
One then finds that $f_{\rm IX}' = (\Gamma_{9}'/\Gamma_{9}^2 -\Gamma_{10}'-\cdots)/G_9^2$.
It turns out that the term associated with $\Gamma_9$ in the parentheses is at least a factor
   of $\sim 7$ smaller than the rest in magnitude.
Therefore $f_{\rm IX}'$ is always negative for the temperature range we examine,
   and hence an anti-phase behavior between $f_{\rm IX}$ and $T$.
Moving on to the next ionic state Fe \MyRoman{10}, we find that $f_{\rm X}'$ is always positive.
This can be understood with Equation~\eqref{eq_EI_fq}, which leads to that 
   $(\ln f_{\rm X})' = (\ln f_{\rm IX})' + (\ln \Gamma_{10})'$.
We find that $(\ln \Gamma_{10})'$ dominates $(\ln f_{\rm IX})'$,
   and hence an $f_{\rm X}'$ that is always positive despite a negative $f_{\rm IX}'$.
From Fe \MyRoman{11} onward, Equation~\eqref{eq_EI_fq} suggests that 
   $f_{q}'$ is positive definite because both $f_{q-1}'$ and $\Gamma_{q}'$ are positive.
And hence the in-phase relationship between $f_{\rm XII}$ and $T$.

When NEI is incorporated, however,
   a time lag exists between the temperature and ionic fraction variations
   given the finite ionization and recombination timescales.
The magnitude of variations in $f$ is also weaker than in the EI case,
   which is true for both ionization states. 
Focusing on the NEI results, one finds that it takes about ${4}$~secs
   for the ionic fractions to settle to a stationary state.
In the first $\sim {4}$~secs, the magnitude of $f_{\rm IX}$ ($f_{\rm XII}$)
   decreases (increases) slightly with time. 
This behavior can be explained by Equation~\eqref{eq_ionic_frac} as follows, 
   which turns out to be rather involved.
To start, the advection term $\vec{v}\cdot\nabla f_q$ turns out to be negligible
   throughout the entire computational domain.
To see this, we note that this term is associated with a frequency $\vec{v}\cdot\nabla$,
   which is dominated by $v_r \partial/\partial r$.
Replacing $\partial/\partial r$ with $1/R$ and noting that $v_r$ reaches up to $30$~km~s$^{-1}$,
   one finds that $\vec{v}\cdot\nabla \sim 0.02$~Hz, which is much smaller than 
   $\omega = 2\pi/P = 1.01$~rad~s$^{-1}$.
This makes $f_q$ effectively local given that the right-hand side (RHS)
   of Equation~\eqref{eq_ionic_frac} involves only the values evaluated at 
   a given location.
We neglect the advection term in the following discussions, 
   and see $f_q$, $N$, $C_q$, and $R_q$ as functions of $t$
   because we are examining a fixed location.
Now define the ionization ($\zeta_{\rm C}$) 
   and recombination  ($\zeta_{\rm R}$) frequencies as
\begin{eqnarray}
&&  \zeta_{{\rm C}, q} = N^{(0)} C_{q}^{(0)}~, \label{freq_ioni} \\
&&  \zeta_{{\rm R}, q} = N^{(0)} R_{q}^{(0)}~, \label{freq_recom} 
\end{eqnarray}
   where the superscript $0$ denotes the values at $t=0$.
Define further that 
\begin{eqnarray}
\Delta g (t) = g (t) - g^{(0)}~,
\end{eqnarray}
   where $g$ denotes $f_q$, $N$, $C_q$, and $R_q$.
Note that $\Delta f_q (t=0) = 0$ by definition.
Equation~\eqref{eq_ionic_frac} then becomes
\footnote{
The terms on the RHS of Equation~\eqref{eq_Deltafq_gen} involve at least 
    one symbol with $\Delta$.
This is because we initiate the solution procedure with the EI solution,
    meaning that $f_{q-1}^{(0)} C_{q-1}^{(0)} = f_{q}^{(0)} R_{q}^{(0)}$.
While choosing this initial condition seems arbitrary,
    the solution procedure needs to be
    initiated at any rate and adopting the EI solution 
    has been a common practice~\citep[e.g.,][]{2010ApJ...722..625K,2013ApJ...773..110S}. 
}
\begin{eqnarray}
\displaystyle  \frac{{\rm d} \Delta f_q}{{\rm d}t}
 = &&  \zeta_{{\rm C}, q-1} \Delta f_{q-1}
          -(\zeta_{{\rm C}, q}+\zeta_{{\rm R}, q})\Delta f_{q}
          +\zeta_{{\rm R}, q+1} \Delta f_{q+1} \nonumber \\
   && +\Delta_{1} + \Delta_{2}+\Delta_{3}~,
     \label{eq_Deltafq_gen}
\end{eqnarray}
   where 
\begin{eqnarray}
\Delta_1 = N^{(0)}\left[
     f_{q-1}^{(0)}\Delta C_{q-1}
      -f_{q}^{(0)}(\Delta C_{q}+\Delta R_{q})
    +f_{q+1}^{(0)}\Delta R_{q+1}  
     \right]~.
\label{eq_Delta1}     
\end{eqnarray}
Furthermore, $\Delta_2$ involves terms like $\Delta C_{q-1}\Delta f_{q-1}$ and $\Delta N \Delta C_{q-1}$,
    while $\Delta_3$ involves such terms as
    $\Delta N \Delta C_{q-1}\Delta f_{q-1}$.    
It turns out that $\Delta_3$ can be safely neglected but the same is not true for $\Delta_2$
    despite that we are actually examining linear waves. 
Nonetheless, we make two simplifications for tractability and see whether the approximate solutions
    are good enough afterwards. 
One is that 
    $\Delta_2$ and $\Delta_3$ can be omitted from Equation~\eqref{eq_Deltafq_gen}.
The other is that $\Delta C$ and $\Delta R$, when seen as functions of $T$, 
    involve only $\Delta T$.
In other words, $\Delta C = C' \Delta T$ and $\Delta R = R' \Delta T$, 
    where $'$ denotes the derivative with respect to $T$ as evaluated 
    at $T = T^{(0)}$.
Note that $\Delta T \propto \sin(\omega t)$.    
As a result, Equation~\eqref{eq_Deltafq_gen} becomes a set of linear ordinary differential equations
    that can be put in matrix form as
\begin{eqnarray}
\displaystyle 
   \frac{{\rm d} \Delta f(t)}{{\rm d} t} = M \Delta f(t) + D \Delta T(t)~,  
   \label{eq_Deltafq_lin_Matrx} 
\end{eqnarray}
    where the column vector $\Delta f(t) \equiv \left[\Delta f_{\rm V}, \Delta f_{\rm VI}, \cdots, \Delta f_{\rm XV}\right]^{\rm T}$.
The constant coefficient matrix $M$ is a tri-diagonal one, for which the non-zero elements
    can be readily recognized from the first row on the RHS of Equation~\eqref{eq_Deltafq_gen}. 
On the other hand, the elements in the constant column vector $D$ read 
\begin{eqnarray}
    D_q = N^{(0)}\left[
     f_{q-1}^{(0)} C_{q-1}'
      -f_{q}^{(0)}(C_{q}'+ R_{q}')
    +f_{q+1}^{(0)} R_{q+1}'  
     \right]~.
\label{eq_Dq}     
\end{eqnarray}
Despite the rather complicated form of $M$ and $D$, Equation~\eqref{eq_Deltafq_lin_Matrx} 
    is in fact a textbook problem.
In short, its solution, subject to the initial condition that $\Delta f_{q} (t=0) = 0$,
    comprises 
    terms that involve either $\exp(\lambda_q t)$ or $\sin (\omega t + \alpha_q)$,
    where $\lambda_q$ represents the eigenvalues of the matrix $M$
    and $\alpha_q$ represents some phase angle.
We find that all values of $\lambda_q$ are real and distinct, with all but one being 
    negative.
Note that one eigenvalue has to be zero because Equation~\eqref{eq_Deltafq_lin_Matrx} 
    guarantees that $\sum_q \Delta f_q = 0$ at all times.
The end result is that as time proceeds, the terms involving $\exp(\lambda_q t)$
    with negative $\lambda_q$ damp out, and the solution becomes sinusoidal. 
The factors in front of $\exp(\lambda_q t)$ are $q$-dependent, and therefore 
    the duration it takes for $\Delta f_q$ to become sinusoidal also depends on $q$.
For instance, the transitory phase for Fe \MyRoman{14} turns out to last for $\sim {12}$~secs.
However, for both Fe \MyRoman{9} and Fe \MyRoman{12}, the transitions to a stationary state both 
    take only about several seconds.
And in the transitory phase, $\Delta f_{\rm IX}$ ($\Delta f_{\rm IX}$) 
    turns out to decrease (increase) slightly.
All these behaviors are in close agreement with the blue curves in Figure~\ref{fig:ion_fraction}.
In fact, despite the two simplifying assumptions behind Equation~\eqref{eq_Deltafq_lin_Matrx},
    its solution is accurate to within ${1.12\%}$ (${10.6\%}$)
    for Fe \MyRoman{9} (Fe \MyRoman{12}).

We now compute the emissivity at each grid point $(r, z)$ at time $t$ via
\begin{eqnarray}
\label{eq_def_eps}
\epsilon = G_{\lambda0} N^2~,
\end{eqnarray}
    where
\begin{eqnarray}
\label{eq_def_G}
   G_{\lambda0} = h\nu_{ij}\cdot0.83\cdot Ab({\rm{Fe}}) f_q \frac{n_jA_{ji}}{N}
\end{eqnarray}
   is the contribution function.
Here $h\nu_{ij}$ is the energy level difference,
   $Ab({\rm{Fe}})$ is the abundance of Fe relative to Hydrogen,
   $f_q$ is once again the ionic fraction of Fe in ionic state $q$,
   $n_j$ is the fraction of Fe in state $q$ lying in level $j$,
   and $A_{ji}$ is the spontaneous transition probability.
We compute $G_{\lambda0}$ with the function \textbf{g\_of\_t} in CHIANTI
   for both Fe \MyRoman{9} 171 \AA\ and Fe \MyRoman{12} 193 \AA.
For the ionic fractions ($f_q$), we consider
   both the EI and NEI values.

Both the line intensities and spectral profiles depend on the LoS.
For convenience, we convert the computed data from the cylindrical
    to a Cartesian grid where the spacing is $30$~km in all three directions.
Note that only $\epsilon$, $T$, $v_r$, and $v_z$ are needed, and
    appropriate interpolation is necessary.
A data hyper-cube in $(x, y, z; t)$ results.
For each LoS, we consider photons emitted from two squares
    of different sizes when projected onto the plane of sky.
Two sizes are considered, one being $30$~km (labeled the ``thin'' beam hereafter)
    and the other $720$~km (or equivalently 1\arcsec, labeled the ``thick'' beam).
For a thin beam,
    we compute the specific intensity ($I$) by integrating $\epsilon$
    with a spacing of $30$~km along the LoS.
On the other hand, we discretize a thick beam into a series of thin beams,
    and compute $I$ by summing up the contributions from all individual thin beams
\footnote{We omit the geometric factor $1/4\pi$ when computing $I$ because
    only the relative variations in $I$ will be examined.
Here by ``relative variations'', we mean $I/I_0$ where $I_0 \equiv I (t=0)$ (see e.g.,
   Figure~\ref{fig:intensity}).
The plane of sky (PoS) becomes different when we switch from one LoS to another. 
For both lines of sight, we make sure that the squares are either 30~km or 720~km 
    across when projected onto the respective PoS.  
It then follows that $I/I_0$ starts from unity when $t=0$ by definition. 
The absolute value of $I_0$ is indeed different for different lines of sight when the square size is fixed,
    or for different square sizes along a given line of sight. 
}.
The spectral profiles are found with the same procedure by integrating
    $\epsilon_\lambda$ at a wavelength $\lambda$ off line center $\lambda_0$.
Following \cite{2016FrASS...3....4V},  $\epsilon_\lambda$ is given by
\begin{eqnarray}
\epsilon_\lambda = \frac{2\sqrt{2\ln2}}{\sqrt{2\pi}\lambda_w}\epsilon
\exp\left(-\frac{4\ln2}{\lambda_w^2}\left(\lambda-\lambda_0\left(1-v_{\rm LoS}/c\right)\right)^2\right)~,
\label{eq_def_eps_lambda}
\end{eqnarray}
where $\lambda_w=(2\sqrt{2\ln2})\lambda_0 (v_{\rm th}/c)$ is the full-width at half-maximum with
    $v_{\rm th}$ ($\propto \sqrt{T}$) being the thermal speed determined by the instantaneous temperature.
Furthermore, $v_{\rm LoS}$ is the velocity projected onto a LoS,
    which in turn is found from the instantaneous flow velocity.
Similar to AvD13, we take $\lambda$ to range from $\lambda_0-0.07$~\AA\ to
     $\lambda_0$+0.07~\AA\ with a spacing of 1.4~m\AA.

\section{Results}
\label{Results}

To start, Figure \ref{fig:intensity} examines
    the temporal evolution of the specific intensities of
	Fe \MyRoman{9} 171 \AA~(the left column)
	and Fe \MyRoman{12} 193 \AA~(right)
	for LoS 1 (the top row) and LoS 2 (bottom).
For the ease of comparison, these intensities have been normalized
    by their values at time $t=0$.
The EI (the red curves) results are shown for comparison with the NEI (blue) results,
    and the effects of different beam sizes are also examined
    with the results for thin (thick) beams shown by the solid (dotted) curves.
Before anything, let us note that
    the difference between any solid curve and its dotted counterpart
    is marginal.
This agrees with previous results
    by \citeauthor{2012A&A...543A..12G}~(\citeyear{2012A&A...543A..12G}, see also AvD13)
    in that a beam size comparable with the half-width
    of the coronal loop is still adequate
    for resolving the sausage mode.
Note that sausage modes are unlikely to be sensitive to the
    fine structuring transverse
    to coronal loops~\citep[e.g.,][]{2007SoPh..246..165P, 2015SoPh..290.2231C}.
Note further that coronal loops typically possess apparent
    widths over a couple of arcsecs~\citep[e.g.,][]{2004ApJ...600..458A, 2007ApJ...662L.119S}.
In what follows we will discuss only the results pertinent to the thin beams
    because a resolution of~1\arcsec\ is readily achievable with modern spectrometers
    like Hinode/EIS~\citep{2007SoPh..243...19C} and IRIS~\citep{2014SoPh..289.2733D}.

Whether or not NEI is considered,
    the intensity variation is consistently stronger
    for LoS 1 than for LoS 2.
This is primarily because LoS 1 samples the portions where
    the density varies in phase,
    whereas LoS 2 samples areas where compression and rarefaction are both present
   (see Figure~\ref{fig:sausage_perturb}c).
In addition, the intensity variation in Fe \MyRoman{12} 193 \AA\
    is consistently stronger than in Fe \MyRoman{9} 171 \AA.
This comes largely from the opposite temperature
    dependence of the contribution functions for the two lines.
While $G$ increases with $T$ for Fe \MyRoman{12} 193 \AA,
    it follows the opposite trend for Fe \MyRoman{9} 171 \AA.
Now that the density $N$ always varies in phase with $T$,
    the product $G_{\lambda} N^2$ and hence $\epsilon$
    possesses a stronger variation for the Fe \MyRoman{12} 193 \AA\ line (see Equation~\ref{eq_def_eps}).

Now move on to the effects of NEI.
Evidently, for the parameters we choose,
   introducing NEI enhances the intensity variation for Fe \MyRoman{9} 171 \AA, whereas the opposite happens for Fe \MyRoman{12} 193 \AA.
This effect is readily seen for LoS 1 (Figs.~\ref{fig:intensity}a and \ref{fig:intensity}b),
   and can also be discerned for LoS 2 (Fig.~\ref{fig:intensity}d).
To understand why NEI impacts the two spectral lines differently,
   we take LoS 1 and examine only the interval between $3$
   and $5$~secs, because LoS 2 and other intervals can be understood in the same way.
Figure~\ref{fig:ion_fraction} indicates that, in this time interval,
   the ionic fraction $f_{\rm IX}$ ($f_{\rm XII}$) is smaller (larger) for EI than for NEI, despite that the overall variations in ionic fractions are consistently stronger when EI
   is assumed.
Given that $G_\lambda$ is proportional to the ionic fraction,
   one finds that $G_{\lambda} N^2$ is smaller (larger) in the EI case for Fe~\MyRoman{9} (Fe~\MyRoman{12}).

Figure~\ref{fig:spec_171} presents the synthesized
    spectral profiles for Fe \MyRoman{9} 171 \AA\
    for both LoS 1(the upper row) and LoS 2 (lower).
Given in the left and middle columns are their temporal evolution
    when EI and NEI are adopted, respectively.
An inspection of these columns indicates that the most obvious difference between 
    the EI and NEI results lies in the temporal variations in the intensity ($I_{\lambda_0}$) 
    attained at the rest wavelength $\lambda_0$ for LoS 1. 
{
While $I_{\lambda 0} (t)$ is enhanced once every half the wave period ($P/2$) for 
    both EI and NEI, the magnitude of the enhancement is nonetheless  different
    when $t$ differs by $P/2$ in the NEI case
    (Figure~\ref{fig:spec_171}b).
}    
Take $t=P/4=1.56$~sec and $t=3P/4=4.66$~sec. 
The values of $I_{\lambda_0}$ are approximately the same for EI
    but show some evident difference for NEI.
This behavior can be understood as follows.
First, at these instants of time, the fluid velocities are zero along LoS 1, 
    which is actually true for the entire computational domain because $\cos(\omega t) = 0$
    (see Equation~\ref{eq_pert_vr} and \ref{eq_pert_vz}).
The exponential term can then be dropped from Equation~\eqref{eq_def_eps_lambda}, and therefore 
    $I_{\lambda 0}$ becomes a LoS integration of $\epsilon_{\lambda0}$
    that is proportional to $\epsilon/\sqrt{T}$.
Now see the contribution function ($G_{\lambda 0}$) and consequently the emissivity ($\epsilon$)
    as functions of electron density ($N$) and temperature ($T$) in view of
    Equations~\eqref{eq_def_G} and \eqref{eq_def_eps}.
Define $G_{T} = \partial G_{\lambda 0}/\partial T$
    and $G_{N} = \partial G_{\lambda 0}/\partial N$, both evaluated
    at the equilibrium values $(N_0, T_0)$.
Define further that $\Delta N = N-N_0$ and $\Delta T = T-T_0$, and recall that 
    $\Delta T/T_0 = (\gamma-1) \Delta N/N_0$.
Despite the specific form of $G_{\lambda 0}$, we find that the variation in $\epsilon_{\lambda 0}$
    is largely determined by first-order perturbations in $N$ and $T$.
In other words, 
\begin{eqnarray}
\displaystyle 
   \frac{\epsilon}{\sqrt{T}} 
   \approx \frac{G_0 N_0^2}{\sqrt{T_0}}
           \left[1+
             \left(\frac{G_N N_0+(\gamma-1)G_T T_0}{G_0}+2-\frac{\gamma-1}{2}
             \right)\frac{\Delta N}{N_0}
           \right]~,
   \label{eq_epslambda_lin}           
\end{eqnarray}
   where $G_0$ is $G_{\lambda 0}$ evaluated at $(N_0, T_0)$.
Now that $\Delta N/N_0 \propto \sin(\omega t)$, one finds that 
   the value that $\epsilon/\sqrt{T}$ attains at $t=P/4$ ($\sin(\omega t) =1$)
   is different from the value at $t=3P/4$ ($\sin(\omega t) =-1$).
Note that while Equation~\eqref{eq_epslambda_lin} pertains only to a fixed location,
   the contribution from the first-order terms survives the LoS integration process.
As a result, in general $I_{\lambda 0}$ at these instants of time should be different 
   in both the EI and NEI cases,
{
   meaning that, strictly speaking, $I_{\lambda 0}$ oscillates 
   at the wave period ($P$) whether or not NEI is considered. 
}
It is just that, for Fe \MyRoman{9} 171 \AA, the difference is not as obvious when EI is adopted,
   and the reason is that $G_N$ is effectively absent in the EI case 
   but plays a substantial role in the NEI case. 

The differences in $I_{\lambda 0}$ notwithstanding, 
   the spectral profiles are remarkably similar in the EI and NEI results.
To quantify this, at each instant of time, 
    we also conduct Gaussian fitting to the instantaneous line profile
    such that the Doppler velocity and width are derived.
These values are presented in the right column as functions of time,
    and we distinguish between the EI (the red curves) and
    the NEI cases (blue).
For both Lines-of-Sight, 
    NEI does not introduce any appreciable difference
    to either the Doppler velocity or width.
This was anticipated by AvD13 on the basis of
    Equation~\eqref{eq_def_eps_lambda} given that
    the ionic fraction enters into discussion only through $\epsilon$, which
    does not affect how $\epsilon_\lambda$ depends on $\lambda$.
However, while this is obvious at any given location, a synthesized
    spectral profile is in fact a LoS integration, meaning that
    the relative contributions from emitting materials
    actually depend on $\epsilon$, which in turn depends on the ionic fraction.
The right column of Figures~\ref{fig:spec_171} is reassuring in the sense that,
    at least for the parameters we choose,
    the spectral profiles of Fe \MyRoman{9} 171~\AA\ can indeed be analyzed without invoking
    the involved NEI effects.
The same can be said for Fe \MyRoman{12} 193~\AA\ (not shown) as far as the effects of NEI
    on the Doppler velocities and widths are concerned.
In fact, the Doppler velocities and widths found with Fe \MyRoman{12} 193~\AA\
    are identical to what we have for Fe \MyRoman{9} 171~\AA.

Compared with NEI, viewing angles play a far more important role
    in determining Doppler velocities and widths.
For LoS 1, the Doppler speed is identically zero (see Figure~\ref{fig:spec_171}c),
    which is expected given that the contributions to LoS 1
    from outward and inward moving fluid parcels cancel out each other.
These bulk motions then contribute to the Doppler broadening,
    which oscillates at half the wave period ($P/2$, Figure~\ref{fig:spec_171}d).
For LoS 2, however, the LoS velocities survive the integration process,
    resulting in a Doppler velocity oscillating at the wave period (Figure~\ref{fig:spec_171}g).
These bulk motions (rather than thermal motions) also contribute to the Doppler broadening,
    which also possesses a period of $P/2$ (Figure~\ref{fig:spec_171}g).

Then what will be the tell-tale signatures of NEI in observations?
The comparison of Figure~\ref{fig:intensity} with \ref{fig:spec_171}
    indicates that LoS 1 does not help for this purpose, because the intensity
    and Doppler width signals oscillate with different periods,
    and a phase-relation analysis is not straightforward.
Considering LoS~2, one finds that Fe \MyRoman{9} 171~\AA\
    is not helpful either because the intensities variations are extremely weak (Fig.~\ref{fig:intensity}c).
For this LoS, however, one may focus on Fe \MyRoman{12} 193~\AA\
    and examine the intensity series
    (Figure~\ref{fig:intensity}d)
    against the Doppler width variations (Figure~\ref{fig:spec_171}h).
For both EI and NEI computations, these two time series possess the same period of $P/2$.
Nonetheless, they are $180^\circ$ out-of-phase provided that Fe is in ionization balance.
On the contrary, for the parameters we choose, a phase difference of $\sim 90^\circ$ is seen between the two time series
    when NEI is incorporated.
We note that while both intensity and Doppler width variations are not that strong,
    they are not undetectable with, say, IRIS
    (see, e.g., Figure~3 in \citeauthor{2016ApJ...823L..16T}~\citeyear{2016ApJ...823L..16T}).

\section{Summary}
\label{summary and conclusion}
This work was motivated by the notion that, for fast sausage modes in coronal loops,
    Iron (Fe) may not be able to maintain ionization balance even for relatively dense loop plasmas.
To address how non-equilibrium ionization (NEI) affects the modulated emissions,
    we plugged the self-consistently derived ionic fractions into the contribution functions
    for both Fe \MyRoman{9} 171~\AA\ and \MyRoman{12} 193~\AA,
    thereby synthesizing both their specific intensities and spectral profiles.
We find that relative to Equilibrium Ionization (EI), NEI plays a far more important role
    in affecting specific intensities than in
    determining Doppler velocities or widths.
We also find that, for the parameters we choose,
    NEI may affect the phase-relation between the intensity variations and those
    in the Doppler widths for Fe \MyRoman{12} 193~\AA.
For lines-of-sight oblique to the loop axis, the two time series possess a phase difference
    of $\sim 90^\circ$ when NEI is incorporated, whereas the phase difference
    is $180^\circ$ when ionization balance is assumed.

\begin{figure}
  \begin{centering}
  \includegraphics[width=0.7\linewidth]{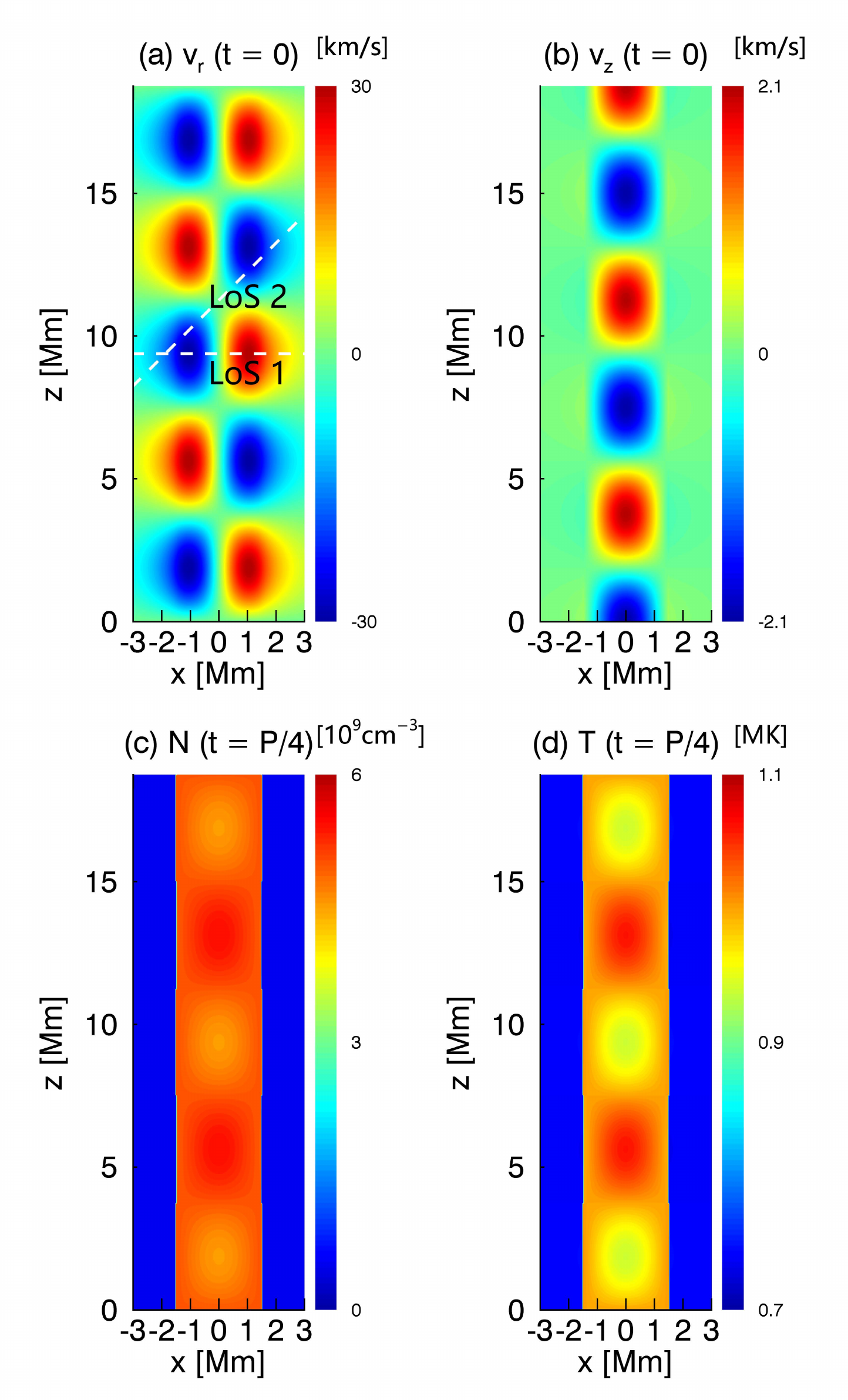}
  \caption{Snapshots of spatial distributions
    of the flow parameters associated with the fast sausage mode.
  Shown here is a cut through the cylinder axis.
  The radial ($v_r$, panel a) and axial ($v_z$, panel b) velocities are for $t=0$,
      while the electron number density ($N$, panel c) and temperature ($T$, panel d)
        are for $t$ being 1/4 the wave period.
  The white dashed lines show the lines-of-sight that this study examines.}
  \label{fig:sausage_perturb}
 \end{centering}
\end{figure}

\begin{figure}
\begin{centering}
 \includegraphics[width=\linewidth]{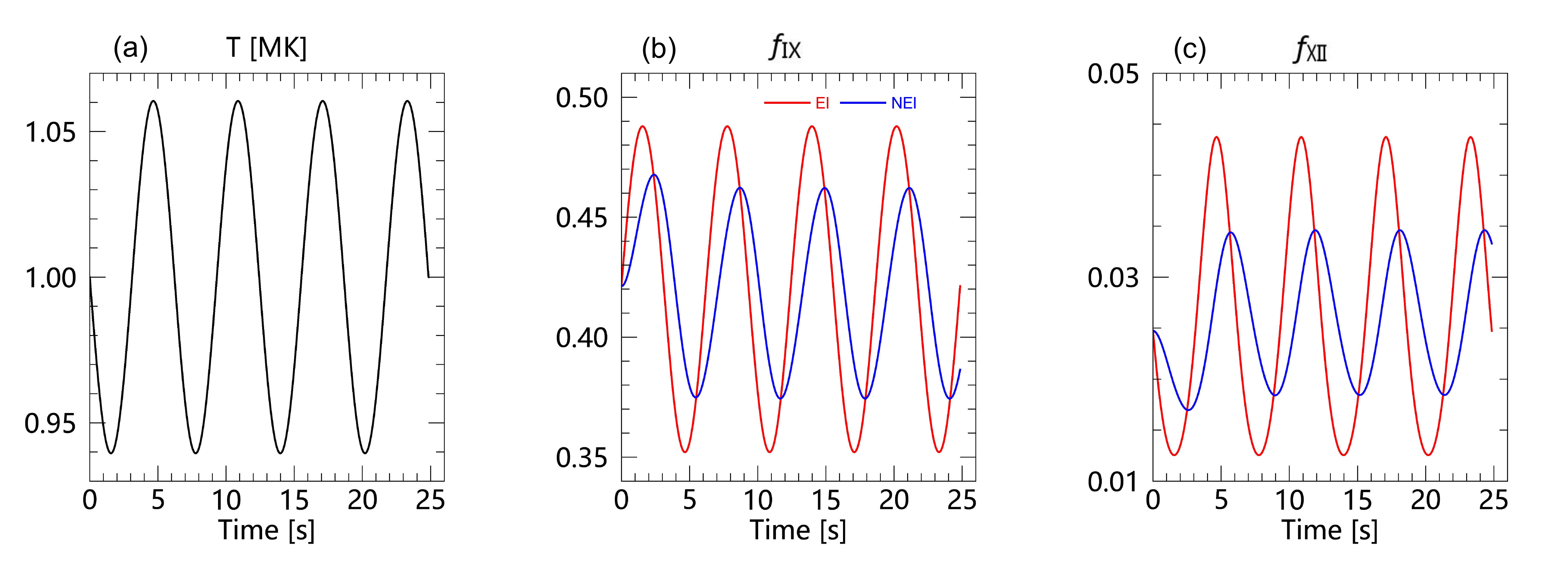}
 \caption{
 Temporal variations of
     (a) the electron temperature,
     (b) the ionic fraction of Fe \MyRoman{9},
     and (c) that of Fe \MyRoman{12}.
 The values are taken at $[r, z] = [0, L/2]$ with $L$ being the cylinder length,
     where the compressibility is the strongest.
 In (b) and (c), the results from the Non-equilibrium-Ionization computation (NEI, the blue curves) are compared with
     the Equilibrium Ionization (EI, red) results.
 }
 \label{fig:ion_fraction}
\end{centering}
\end{figure}

\begin{figure}
  \begin{centering}
  \includegraphics[width=\linewidth]{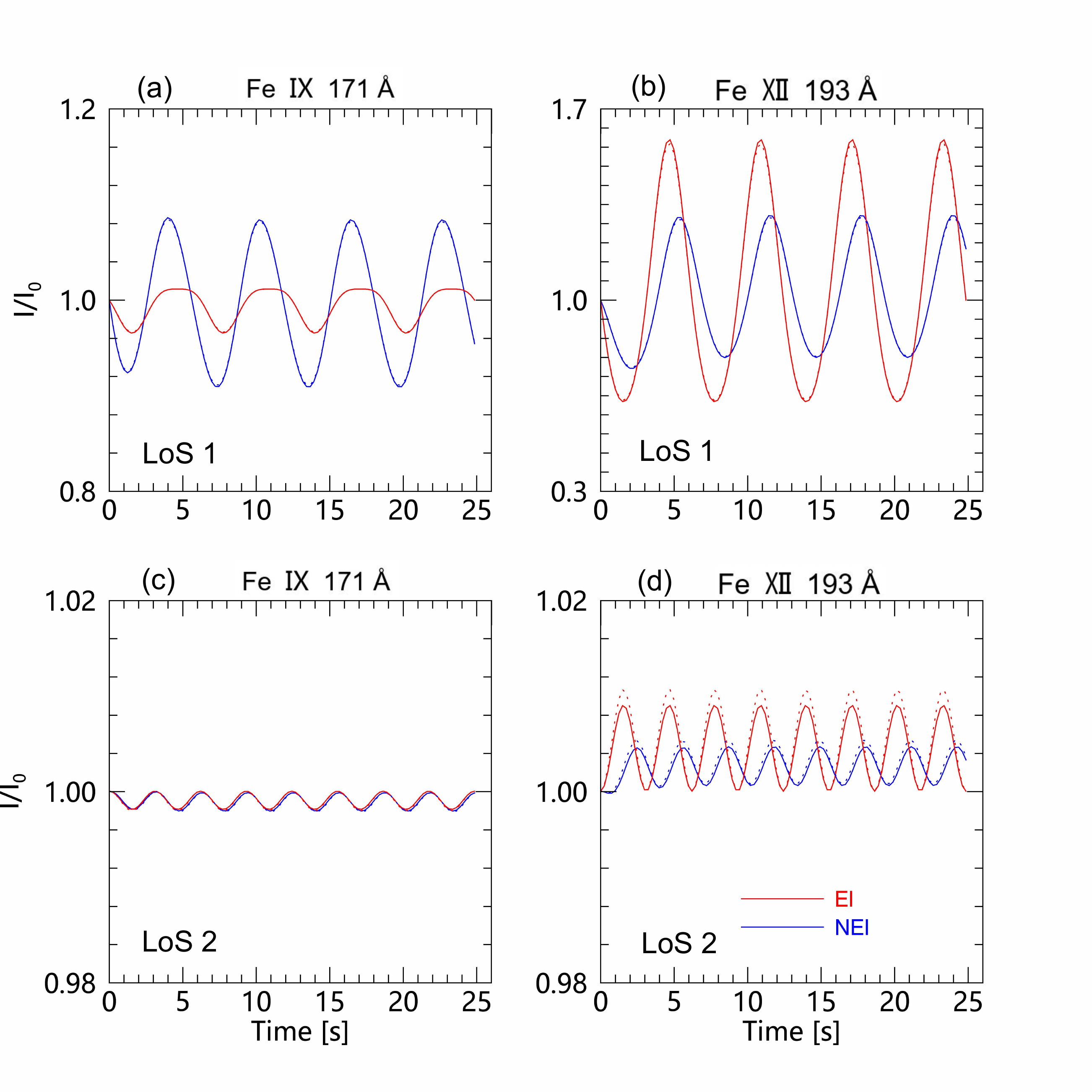}
   \caption{Synthesized specific intensities normalized by their values at $t=0$
     for Fe \MyRoman{9} 171 \AA~(the left column)
       and Fe \MyRoman{12} 193 \AA (right).
   Two LoS, 1 and 2, are examined in the upper and lower rows, respectively.
   The red and blue lines represent, respectively, the Equilibrium Ionization (EI) and
       Non-equilibrium Ionization (NEI) results.
   Furthermore, the solid (dotted) lines are for a beam $30$~km ($720$~km) across when projected
       onto the plane-of-sky, see text for details.}
 \label{fig:intensity}
\end{centering}
\end{figure}

\begin{figure}
  \begin{centering}
  \includegraphics[width=0.8\linewidth]{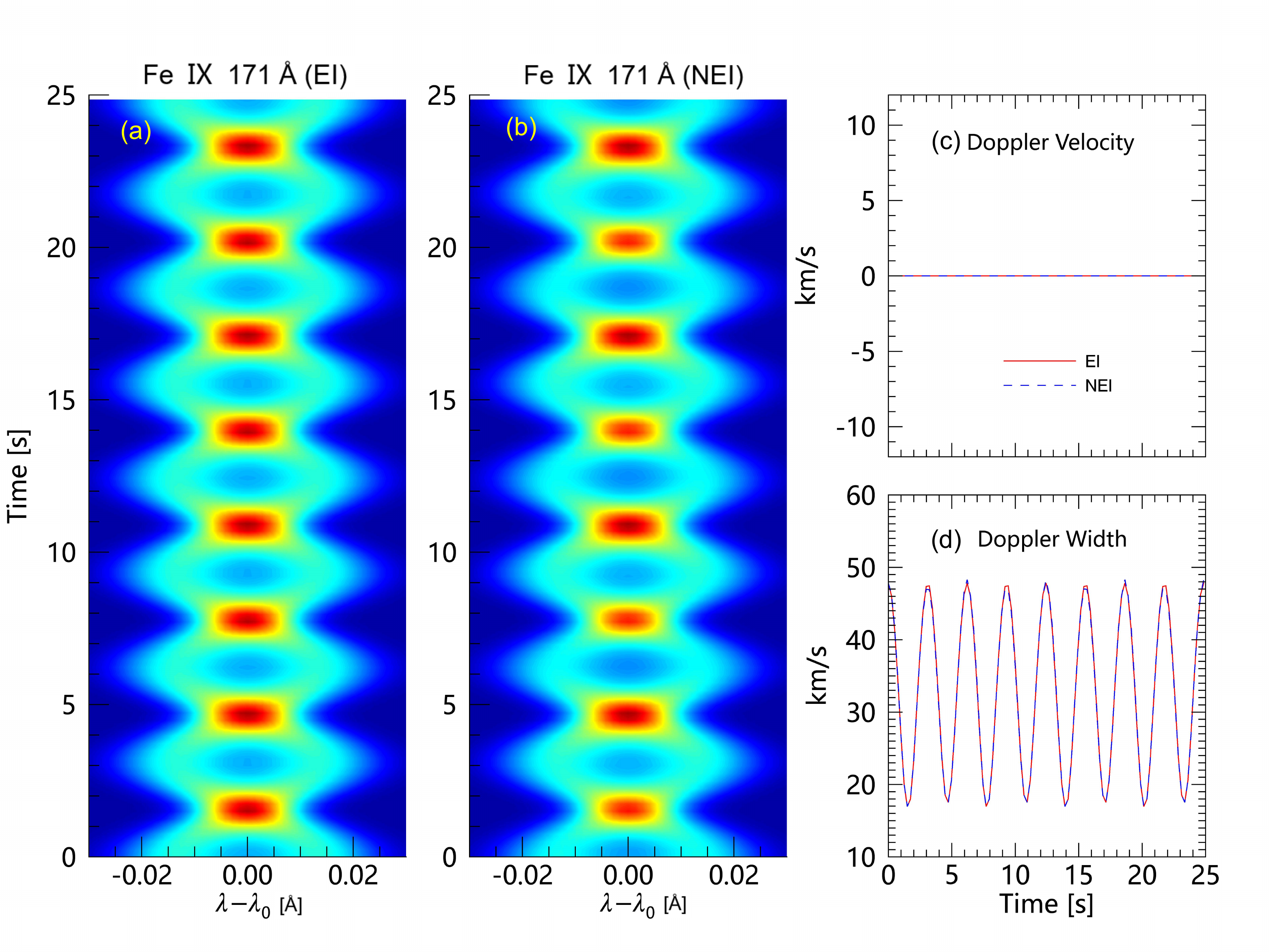}
  \includegraphics[width=0.8\linewidth]{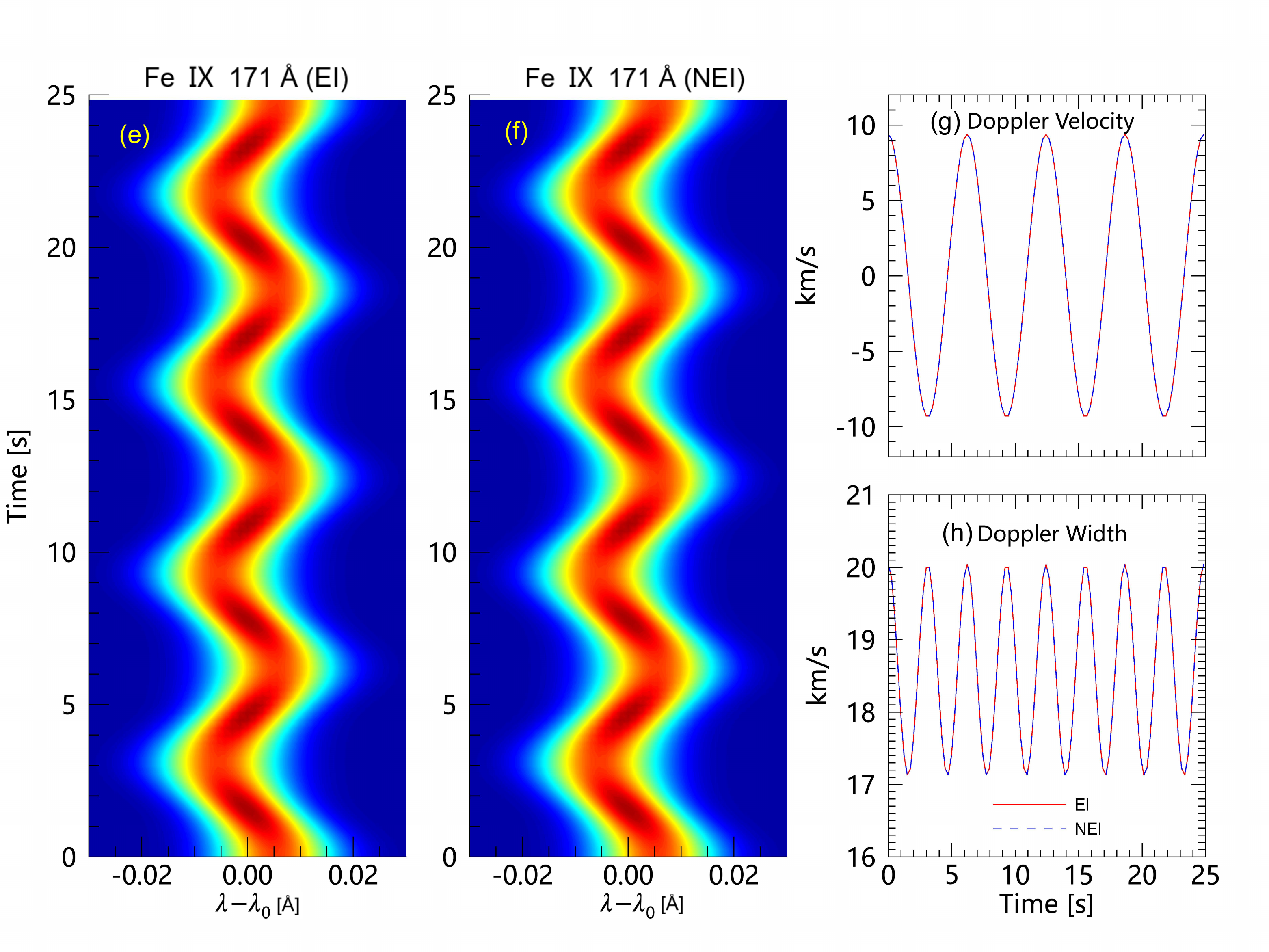}
  \caption{Spectral properties for Fe \MyRoman{9} 171 \AA\ along LoS 1 (the upper row) and 2 (lower).
  The left and middle columns show the spectral profiles
      when Equilibrium Ionization (EI) and Non-equilibrium Ionization (NEI) are adopted, respectively.
  In the right column, the Doppler velocity and width are shown for both the EI
      (the red solid lines) and NEI (blue dashed) cases.
  }
 \label{fig:spec_171}
\end{centering}
\end{figure}

{
Before closing, let us discuss some limitations of the present study and hence the ways to move forward.
With a length-to-radius ratio $L/R = 12.5$ and a loop radius $R= 1.5$~Mm,
    one finds a loop length $L \approx 19 $~Mm. 
While this loop length is not unrealistic, it is nonetheless on the low side of the observed range of the lengths of the EUV loops
    \citep[see e.g., Figure 1 in][]{2007ApJ...662L.119S}.
Furthermore, while there is observational evidence showing the possible existence of the first longitudinal harmonic
    of FSMs in flare loops~\citep[e.g.,][]{2003A&A...412L...7N, 2005A&A...439..727M, 2008MNRAS.388.1899S},
    the observations of higher harmonics in EUV-emitting active region loops have yet to be found. 
The reason for us to choose a relatively short loop and a higher harmonic is to make sure
    that the fast sausage mode (FSM) is trapped. 
As is well-known, FSMs are trapped only when the dimensionless axial wavenumber ($kR$)
    exceeds some critical value ($(kR)_{\rm cutoff}$)~\citep[e.g.,][]{1983SoPh...88..179E, 1986SoPh..103..277C, 2007AstL...33..706K}. 
Let $n$ denote the axial harmonic number with $n=0$ representing the fundamental mode by convention. 
The dimensionless axial wavenumber $kR$ is then $(n+1) \pi R/L$. 
This means that trapped modes are allowed only when $n$ is sufficiently large and/or the loop is sufficiently short. 
When the plasma beta is small, $(kR)_{\rm cutoff}$ is largely determined by the density contrast
    between the loop and its ambient~\citep[e.g.,][Equation~5]{2007AstL...33..706K}. 
We find that $(kR)_{\rm cutoff} = 0.793$ for the physical parameters we choose, meaning that 
    $n$ needs to be at least three for $kR$ to exceed the critical value for the examined length-to-radius ratio.
The fourth harmonic ($n=4$) is nonetheless chosen, largely compatible with previous forward modeling studies
    by \citet{2012A&A...543A..12G} and \citet{2013A&A...555A..74A}.
The reason for us to stick to the trapped modes is that we would like
    to avoid further complications associated with the temporal attenuation of the leaky modes. 
While an eigen-mode analysis is equally possible for both the trapped and leaky modes, 
    the analytically derived eigenfunctions for the leaky ones
    diverge exponentially in the ambient corona~\citep[see e.g.,][Equation~3.1]{1986SoPh..103..277C}. 
The end result is that, while the periods and damping rates are accurately captured by the eigen-mode analysis, 
    the eigen-functions for the leaky modes cannot fully describe a system experiencing sausage mode oscillations. 
Rather, the temporal evolution of the system should be examined from the initial-value-problem perspective
    by using a largely numerical approach~\citep[e.g., appendices in][]{2016SoPh..291..877G,2016ApJ...833..114C}. 
Consequently the emission properties should be computed with the numerically simulated data. 
To make our computations as simple as possible, we choose to work with the trapped modes, 
    for which the analytically derived eigen-functions can fully describe a loop oscillating in an eigen-mode. 
    
Having said that, our results can still find applications even to fundamental modes. 
Firstly, let us consider the case where the fundamental modes are trapped, as would be expected for short and dense 
    flaring loops. 
Physically speaking, the spatial structures of the perturbations associated with higher longitudinal harmonics
    are just a repetition of those associated with the fundamental mode
    (with possible reversal of signs, see Fig.~\ref{fig:sausage_perturb}). 
The consequence is that, at sufficiently high spatial resolution, 
    the emission properties for the fundamental mode will be close to our results
    when one adopts a line of sight that passes through an anti-node. 
This point was also recognized by \citet{2013A&A...555A..74A}, and employed by~\citet{2016ApJ...823L..16T}
    to interpret their IRIS measurements. 
We are currently conducting a study tailored to this latter work on the Fe~\MyRoman{21} 1354~\AA\ emissions modulated by 
    a fundamental FSM. 
Secondly, the Non-equilibrium Ionization (NEI) effects
    are expected to be important for fundamental modes even if they are leaky for typical EUV loops. 
This is because the NEI effects will show up as long as the wave period is not too long when compared
    with the ionization and recombination timescales. 
For the fundamental mode, the period ($P$) is still largely determined by the transverse fast time, 
    and is approximately $(R/v_{\rm Ae}) [2\pi/(kR)_{\rm cutoff}]  \sim 7.1$~secs
    for the loop examined in this manuscript. 
This value is rather close to the period of the higher harmonic we examined, for which $P\sim 6.2$~secs. 
And therefore the deviation of the NEI results from the EI ones are expected. 
A study on the NEI effects on the EUV emissions associated with a leaky fundamental mode is also underway.
}

\acknowledgments
{We thank the referee for his/her constructive comments.}
We thank Drs.~Patrick Antolin and Valery Nakariakov for helpful discussions.
This work is supported by
    the National Natural Science Foundation of China (41474149, 41604145, 41674172, 11761141002).
TVD is supported by the GOA-2015-014 (KU~Leuven) and the European Research Council (ERC) 
    under the European Union's Horizon 2020 research and innovation 
    programme (grant agreement No 724326).
ZH is also supported by the Young Scholar Program of Shandong University Weihai (2017WHWLJH07).
This work is also supported by the Open Research Program of
    the Key Laboratory of Solar Activity of National Astronomical Observatories of China
    (BL: KLSA201801).
CHIANTI is a collaborative project involving George Mason University, 
    the University of Michigan (USA) and the University of Cambridge (UK).

\bibliographystyle{aasjournal}
\bibliography{NEI_saus}



\end{document}